\documentclass[11pt]{article}

\usepackage{graphicx} 
\usepackage{amssymb,amsmath,amsthm,csquotes,mathtools,amssymb}
\usepackage{bbm}
\usepackage[backend=biber,style=authoryear-comp,citestyle=authoryear-comp,maxnames=1]{biblatex}
\usepackage{mathtools}
\usepackage{csquotes}
\usepackage{amsthm}
\usepackage{amsmath}
\usepackage{mathabx}
\usepackage{dsfont}
\usepackage{enumerate}
\usepackage{biblatex}
\usepackage{hyperref}
\usepackage{physics}
\usepackage{multirow,array}
\usepackage{xcolor}

\usepackage{subcaption}
\usepackage[margin=1in]{geometry}
\usepackage{tikz}
\usetikzlibrary{positioning, arrows.meta}

\title{Quantum Primitive for Output-Hiding Function Sharing: QKD and Joint Computation Applications}

\author{Olivia R. Hartzell\thanks{This work is the subject of U.S. Provisional Patent Application no. 64/045,601.\\
\copyright 2026 Olivia Hartzell. This work is licensed under a Creative Commons Attribution-NonCommercial-NoDerivatives 4.0 International License (CC BY-NC-ND 4.0) available at https://creativecommons.org/licenses/by-nc-nd/4.0/. Any commercial reuse or creation of derivative works without the express written consent of the author is prohibited.}}

\date{\today}

\begin{document}
\maketitle

\begin{abstract}
Applications of the proposed primitive: \textit{Quantum Primitive for Output-Hiding Function Sharing} are discussed for secure quantum communications and computation protocols. In particular, QKD applications provide notable enhanced security and efficiency properties relative to status-quo protocols. Additionally, we provide examples when parties may wish to encode \textit{joint} functions, or decisions, which remain information-theoretically hidden from external parties and those internal to the quantum system, without additional private keys, hidden randomness, or classical communication. In particular these applications may be useful in domains such as; financial transactions, joint signaling or coordination decisions, and navigation systems, among others. 
\end{abstract}

\section{Introduction}
We discuss applications of the primitive described in detail in \textit{Quantum Primitive for Output-Hiding Function Sharing} (\cite{hartzell_quantum}).

\subsection{Quantum Communications and QKD}
The primitive provides a process for quantum communications, including quantum key distribution (QKD), that differs notably from existing protocols through both enhanced security and efficiency properties.

First, in standard Measurement Device Independent (MDI)-QKD implementations (\cite{Lo2012MDI}) that use \cite{BennettBrassard1984} states, a malicious measurement referee may learn private information -- or a shared function value $s$-- with seventy-five percent probability on any attack, even when state preparation is trusted. In this protocols, and other entanglement-based protocols (\cite{ekert1991qkd}), the resulting keys are secure conditional on attack due to significant post-processing that obfuscates any information the referee learns during the protocol. 

By contrast, in the proposed primitive, security of the shared function value arises from the physics of the quantum states used to generate shared information -- rather than requiring significant privacy amplification to eliminate information gained during function computation. Importantly, an adversary's likelihood of correctly determining the shared function values is \textit{decoupled} from her likelihood of generating an error. In particular, her best prepare-and-measure strategies may yield zero error, and may not obtain information over function values better than no attack. In this sense, the proposed primitive generates \textit{ideal} outputs, in the presence of prepare-and-measure attacks described in detail in \cite{hartzell_quantum}. 

In some of the proposed protocols, parties may detect errors \textit{during} the protocol itself. Overall, the example protocols in section \ref{sec:qkd} benefit from improved efficiency properties relative to Measurement Device Independent (MDI)-QKD style protocols (\cite{Lo2012MDI}), doubling or in some cases, quadrupling, the effective key rate in the ideal execution environment. 

Additionally, key parameters, including the probability that parties record a shared function value for any given run of the primitive are \textit{tunable}, and may be traded off with any attacker's ability to gain information, including learning about \textit{when} a function value gets recorded. Finally, the primitive eliminates the basis sifting process in \cite{BennettBrassard1984}-style protocols, which requires authenticated classical channels.

\subsection{Joint Computations}
While QKD protocols only permit parties to share a single bit of information, ($s(x_a, x_b) = x_a$), the proposed primitive permits parties to share functions that non-degenerately depend on both of their local inputs $(x_a, x_b)$. The security properties of the primitive therefore allows for parties to generate joint computations or decisions, using the single measurement outcome variable alone. In particular, these resulting function values $s(x_a, x_b)$ satisfy all of the security guarantees discussed in detail in \cite{hartzell_quantum}. 

We remind the reader that the primitive bears a fundamentally different information-sharing property than secure multi-party computation (MPC). In the proposed setting, each party \textit{learns} the others' local information, while their output $s(x_a, x_b)$ remains information-theoretically hidden. This structure contrasts sharply with classical and quantum secure function evaluation protocols (\cite{Yao1982, GoldreichMicaliWigderson1987, CrepeauGottesmanSmith2002}), including quantum secure multi-party summation protocols (\cite{Shi2016}, \cite{Lu2024}, \cite{wu2023quantum}, \cite{yi2021quantum}), in which the parties wish to share a joint function value, while \textit{not} revealing their inputs to one another. In particular, the resulting function value may be public.  
By contrast, in the proposed setting, generating these shared values \textit{does not} require any additional resources, including keys generated by pre-emptive QKD protocols.

Several example implementations and their applications are discussed in section \ref{sec:joint}. All examples that follow correspond to the described primitive and the example implementations in \cite{hartzell_quantum}. 

\section{Quantum Key Distribution Implementations}\label{sec:qkd}

Let $P_{keep}$ denote the key retention rate, or the probability that $\tilde{s} \neq \emptyset$, and $P_{error}$ denote the probability that the protocol generates an error, $s_a \neq s_b$.

In a QKD implementation, parties may choose a recording strategy of $\tilde{s} \in \{0, 1, \emptyset\}$, such as one depicted in Figure  \ref{fig:game_matrix_w_code}. Such an implementation permits parties to obscure \textit{when} a value $s(x_a, x_b)$ gets recorded. In this setting, $P_{keep} = \frac{1}{2}$, twice as large as that of standard MDI-QKD. 

\begin{figure}[ht]
    \centering
    \[
    \begin{array}{c|cccc}
       & U_b & U_b' & U_b'' & U_b''' \\ \hline
    U_a & (\ket{\phi_{00}}, {0}) & (\ket{\phi_{11}}, {1}) & (\ket{\phi_{10}}, {0}) & (\ket{\phi_{01}}, {1})\\
    U_a' & (\ket{\phi_{11}}, {0}) & (\ket{\phi_{00}}, {1}) & (\ket{\phi_{01}}, {0}) & (\ket{\phi_{10}}, {1}) \\
    U_a'' & \ket{\phi_{10}} & \ket{\phi_{01}} & \ket{\phi_{00}} & \ket{\phi_{11}} \\
    U_a''' & \ket{\phi_{01}} & \ket{\phi_{10}} & \ket{\phi_{11}} & \ket{\phi_{00}} \\
    \end{array}
    \]
    \caption{Measurement outcome and recording strategy pairs, given corresponding joint unitaries, where rows correspond to Alice's unitaries and columns correspond to Bob's. Each measurement outcome denoted in an entry occurs with probability one under corresponding joint unitaries. Entries without a recorded value correspond with no recorded value. }
    \label{fig:game_matrix_w_code}
\end{figure}

Other implementations may be adopted such that $P_{keep} = 1$, four times as large as that of standard MDI-QKD, for example in Figure \ref{fig:game_matrix_w_code_keep1}. In this setting, Bob may learn of errors generated \textit{during} the protocol when the measurer Eve makes announcements $j$ that are \textit{infeasible} under his chosen unitary.

In all settings, $P(\hat{s} = s | \tilde{Z}, j) = \frac{1}{2}$, as discussed in detail in \cite{hartzell_quantum}, which importantly does \textit{not} hold in standard \cite{BennettBrassard1984}-type protocols with outsourced measurements, whose security against adversaries requires significant post-processing to obfuscate attackers' knowledge gained \textit{during} key generation. A further discussion of these differences and secure key rates is in Section \ref{ref:secure_function_rates}. 

\begin{figure}[ht]
    \centering
    \[
    \begin{array}{c|cccc}
       & U_b & U_b' & U_b'' & U_b''' \\ \hline
    U_a & (\ket{\phi_{00}}, {0}) & (\ket{\phi_{11}}, {1}) & (\ket{\phi_{10}}, {0}) & (\ket{\phi_{01}}, {1})\\
    U_a' & (\ket{\phi_{11}}, {0}) & (\ket{\phi_{00}}, {1}) & (\ket{\phi_{01}}, {0}) & (\ket{\phi_{10}}, {1}) \\
    \end{array}
    \]
    \caption{Measurement outcome and recording strategy pairs, given corresponding joint unitaries, where rows correspond to Alice's unitaries and columns correspond to Bob's. Each measurement outcome denoted in an entry occurs with probability one under corresponding joint unitaries. Entries without a recorded value correspond with no recorded value. }
    \label{fig:game_matrix_w_code_keep1}
\end{figure}

Additionally, we note that the basis sifting procedure is eliminated by the joint measurement event, which reveals inputs between the parties' without the need for additional classical communication. Finally, we remind the reader that adversarial Eve and Charley's attack that maximizes their probability of $\{\hat{s} = s\}$ subject to minimizing the likelihood of error, $s_a \neq s_b$, permits the parties to learn $s$ with likelihood no better than random, and cause no error. In this sense, the proposed primitive \textit{decouples} adversarial success from $P_{error}$. See a discussion in \cite{hartzell_quantum}.

\subsubsection{QKD 2-Bit String Procedure}
For concreteness, we provide a brief step-by-step illustration of how the above setting along with the recording strategy depicted in Figure \ref{fig:game_matrix_w_code} may implement a 2-bit string used in a QKD implementation. \\

\textbf{Iteration 1: $\tilde{s_1}$}\\
\begin{enumerate}
    \item Alice and Bob receive their relevant qubits $\mathcal{J} \ket{00}$. 
    \item Alice's local randomization selects $U_a$, while Bob's local randomization selects $U_b$. They apply these unitaries
    
\[
U_a = \frac{1}{\sqrt{2}}
\begin{bmatrix}
1 & e^{i \pi/4} \\
- e^{-i \pi/4} & 1
\end{bmatrix}, \quad
U_b = \frac{1}{\sqrt{2}}
\begin{bmatrix}
1 & e^{i \pi/4} \\
- e^{-i \pi/4} & 1
\end{bmatrix}
\]

yielding the quantum state
$$(U_a \otimes U_b) \mathcal{J} \ket{00} 
= \frac{1}{\sqrt{2}} \ket{00} - \frac{i}{\sqrt{2}} \ket{11} 
= \ket{\phi_{00}} $$
which is sent to Eve for measurement. 

    \item Eve measures by $\mathcal{J}^{\dagger}(U_a \otimes U_b) \mathcal{J} \ket{00} = \ket{00}$, measures the computational basis state $\ket{00}$, and announces measurement outcome $\ket{\phi_{00}}$. 

    \item Alice and Bob record $\tilde{s_1}$ according to the strategy prescribed in Figure \ref{fig:game_matrix_w_code}. Since Alice chose $U_a$, and heard measurement announcement, $\ket{\phi_{00}}$, she records $\tilde{s_1} = 0$. Since Bob chose $U_b$, and heard measurement announcement, $\ket{\phi_{00}}$, he records $\tilde{s_1} = 0$.
    
\end{enumerate}

\textbf{Iteration 2: $\tilde{s_2}$}\\
\begin{enumerate}
    \item Alice and Bob receive their relevant qubits $\mathcal{J} \ket{00}$. 
    \item Alice's local randomization selects $U_a''$, while Bob's local randomization selects $U_b'$. They apply these unitaries
    
\[
U_a'' = \frac{1}{\sqrt{2}}
\begin{bmatrix}
e^{i 5\pi/4} & e^{i \pi/2} \\
- e^{-i \pi/2} & e^{-i 5\pi/4}
\end{bmatrix}, \quad
U_b' = \frac{1}{\sqrt{2}}
\begin{bmatrix}
e^{i \pi/2} & e^{i 3\pi/4} \\
- e^{-i 3\pi/4} & e^{-i \pi/2}
\end{bmatrix}
\]

yielding the quantum state
$$(U_a'' \otimes U_b') \mathcal{J} \ket{00} = \frac{1}{\sqrt{2}} \ket{01} - \frac{i}{\sqrt{2}} \ket{10} 
= \ket{\phi_{01}} $$
which is sent to Eve for measurement. 

    \item Eve measures by $\mathcal{J}^{\dagger}(U_a'' \otimes U_b') \mathcal{J} \ket{00} = \ket{01}$, measures the computational basis state $\ket{01}$, and announces measurement outcome $\ket{\phi_{01}}$. 

    \item Alice and Bob record $\tilde{s_2}$ according to the strategy prescribed in Figure \ref{fig:game_matrix_w_code}. Since Alice chose $U_a''$, and heard measurement announcement, $\ket{\phi_{01}}$, she records no value, e.g. $\tilde{s_2} = \emptyset$. Since Bob chose $U_b'$, and heard measurement announcement, $\ket{\phi_{01}}$, he records no value. 

\end{enumerate}    

\textbf{Iteration 3: $\tilde{s_3}$}\\
\begin{enumerate}
    \item Alice and Bob receive their relevant qubits $\mathcal{J} \ket{00}$. 
    \item Alice's local randomization selects $U_a$, while Bob's local randomization selects $U_b'$. They apply these unitaries
    
\[
U_a = \frac{1}{\sqrt{2}}
\begin{bmatrix}
1 & e^{i \pi/4} \\
- e^{-i \pi/4} & 1
\end{bmatrix}, \quad
U_b' = \frac{1}{\sqrt{2}}
\begin{bmatrix}
e^{i \pi/2} & e^{i 3\pi/4} \\
- e^{-i 3\pi/4} & e^{-i \pi/2}
\end{bmatrix}
\]

yielding the quantum state
$$(U_a \otimes U_b') \mathcal{J} \ket{00} = \frac{1}{\sqrt{2}} \ket{11} - \frac{i}{\sqrt{2}} \ket{00} 
= \ket{\phi_{11}} $$
which is sent to Eve for measurement. 

    \item Eve measures by $\mathcal{J}^{\dagger}(U_a \otimes U_b') \mathcal{J} \ket{00} = \ket{11}$, measures the computational basis state $\ket{11}$, and announces measurement outcome $\ket{\phi_{11}}$. 

    \item Alice and Bob record $\tilde{s_3}$ according to the strategy prescribed in Figure \ref{fig:game_matrix_w_code}. Since Alice chose $U_a$, and heard measurement announcement, $\ket{\phi_{11}}$, she records $\tilde{s_3} = 1$. Since Bob chose $U_b'$, and heard measurement announcement, $\ket{\phi_{11}}$, he records $\tilde{s_3} = 1$. 

\end{enumerate}    

The resulting 2-Bit string is therefore: $S = \tilde{s_1}, \tilde{s_3} = 0, 1$.

\section{Joint Computations and Decisions}\label{sec:joint}
We next provide examples of how the proposed primitive additionally permits the parties to share and compute functions $s(x_a, x_b)$, which depend on their \textit{joint information} -- a capability not permitted by QKD alone. \\

This primitive may be adopted for function evaluation problems similar in spirit to the millionaire's problem (\cite{Yao1982}), but instead where parties learn each other's input values while the shared decision $s$ remains hidden from external parties. For example, $a, b$ wish to compute 

\[
s(x_a,x_b) =
\begin{cases}
1, & \text{if } x_a > x_b, \\[6pt]
0, & \text{otherwise.}
\end{cases}
\]
where $x_a$ may take one of two discrete values or ranges, and $x_b$ may take on four. Such a construction may be particularly useful for bi-lateral trade settings, such as in financial markets, where local unitaries map to individual valuations for a good, and $s$ may denote a decision to trade.

\begin{figure}[ht]
    \centering
    \[
    \begin{array}{c|cccc}
       & U_b & U_b' & U_b'' & U_b''' \\ \hline
    U_a & (\ket{\phi_{00}} , {1}) & (\ket{\phi_{11}}, {0}) & (\ket{\phi_{10}}, {0}) & (\ket{\phi_{01}}, {0})\\
    U_a'' & (\ket{\phi_{10}}, {1}) & (\ket{\phi_{01}}, {1}) & (\ket{\phi_{00}}, {0}) & (\ket{\phi_{11}}, {1}) \\
    \end{array}
    \]
    \caption{Measurement outcome and recording strategy pairs, given corresponding joint unitaries, where rows correspond to Alice's unitaries and columns correspond to Bob's. Each measurement outcome denoted in an entry occurs with probability one under corresponding joint unitaries.}
    \label{fig:millionare_problem}
\end{figure}



Parties may also compute $s = x_a + 2 x_b \: \text{mod 4}$ as shown in Figure \ref{fig:xa+2xb mod 4}, or $s = x_a - 2x_b \: \text{mod 4}$. For these functions which permit parties to compute the sum or difference over their inputs, may be useful for applications such as coordinated navigation decisions, where $x_i$ may map to parties' coordinates, and $s \in \{0, 1, 2, 3\}$ may denote a joint navigation direction such as $\{N, S, E, W\}$.

\begin{figure}[ht]
    \centering
    \[
    \begin{array}{c|cccc}
       & U_b & U_b' & U_b'' & U_b''' \\ \hline
     U_a & (\ket{\phi_{00}}, {3}) & (\ket{\phi_{11}}, {1}) & (\ket{\phi_{10}}, {3}) & (\ket{\phi_{01}}, {1})\\
    U_a' & (\ket{\phi_{11}}, {0}) & (\ket{\phi_{00}}, {2}) & (\ket{\phi_{01}}, {0}) & (\ket{\phi_{10}}, {2}) \\
    U_a'' & (\ket{\phi_{10}}, 1) & (\ket{\phi_{01}}, 3) & (\ket{\phi_{00}}, 1) & (\ket{\phi_{11}}, 3) \\
    U_a''' & (\ket{\phi_{01}}, 2) & (\ket{\phi_{10}}, 0) & (\ket{\phi_{11}}, 2) & (\ket{\phi_{00}}, 0) \\
    \end{array}
    \]
    \caption{Measurement outcome and recording strategy pairs, given corresponding joint unitaries, where rows correspond to Alice's unitaries and columns correspond to Bob's. Each measurement outcome denoted in an entry occurs with probability one under corresponding joint unitaries. Entries without a recorded value correspond with no recorded value. Unitaries correspond to input values as follows, $1 \iff U_i, 2 \iff U_i', 3 \iff U_i'', 4 \iff U_i'''$. Each recording strategy corresponds with $x_a + 2 x_b \: \text{mod 4}$.}
    \label{fig:xa+2xb mod 4}
\end{figure}


Given that the primitive permits parties to each mutually share two bits of information, implementations such as those depicted in Figure \ref{fig:xor} may be useful for parties' to compare their preferences on two independent inputs, and therefore may be useful for strategic signaling, voting, and DNA matching. For example, $s = x_a \oplus x_b, x_i \in \{00, 01, 10, 11\}$.

\begin{figure}[ht]
    \centering
    \[
    \begin{array}{c|cccc}
       & U_b & U_b'' & U_b''' & U_b' \\ \hline
    U_a & (\ket{\phi_{00}}, 00) & (\ket{\phi_{10}}, 01) & (\ket{\phi_{01}}, 10) & (\ket{\phi_{11}}, 11)\\
    U_a' & (\ket{\phi_{11}}, 01) & (\ket{\phi_{01}}, 00) & (\ket{\phi_{10}}, 11) & (\ket{\phi_{00}}, 10) \\
    U_a'' & (\ket{\phi_{10}}, 10) & (\ket{\phi_{00}}, 11) & (\ket{\phi_{11}}, 00) & (\ket{\phi_{01}}, 01) \\
    U_a''' & (\ket{\phi_{01}}, 11) & (\ket{\phi_{11}}, 10) & (\ket{\phi_{00}}, 01) & (\ket{\phi_{10}}, 00) \\
    \end{array}
    \]
    \caption{Measurement outcome and recording strategy pairs, given corresponding joint unitaries, where rows correspond to Alice's unitaries and columns correspond to Bob's. Each measurement outcome denoted in an entry occurs with probability one under corresponding joint unitaries. Entries without a recorded value correspond with no recorded value. Unitaries correspond to input values as follows, $00 \iff U_a, 01 \iff U_a', 10 \iff U_a'', 11 \iff U_a''', 00 \iff U_b, 11 \iff U_b', 01 \iff U_b'', 10 \iff U_b'''$. Each recording strategy corresponds with $x_a \oplus x_b $.}
    \label{fig:xor}
\end{figure}

Figure \ref{fig:nonlinear} depicts an implementation where $s = 2x_a * x_b + 2 x_a + x_b \: \text{mod 4}, x_i \in \{1, 2, 3, 4\}$. Such implementations may be useful for computing non-linear functions of inputs, where $s$ may be the output of a sensitive medical computation, for example. 
\begin{figure}[ht]
    \centering
    \[
    \begin{array}{c|cccc}
       & U_b & U_b' & U_b'' & U_b''' \\ \hline
     U_a & (\ket{\phi_{00}}, {1}) & (\ket{\phi_{11}}, {1}) & (\ket{\phi_{10}}, {1}) & (\ket{\phi_{01}}, {1})\\
    U_a' & (\ket{\phi_{11}}, {0}) & (\ket{\phi_{00}}, {2}) & (\ket{\phi_{01}}, {0}) & (\ket{\phi_{10}}, {2}) \\
    U_a'' & (\ket{\phi_{10}}, 3) & (\ket{\phi_{01}}, 3) & (\ket{\phi_{00}}, 3) & (\ket{\phi_{11}}, 3) \\
    U_a''' & (\ket{\phi_{01}}, 2) & (\ket{\phi_{10}}, 0) & (\ket{\phi_{11}}, 2) & (\ket{\phi_{00}}, 0) \\
    \end{array}
    \]
    \caption{Measurement outcome and recording strategy pairs, given corresponding joint unitaries, where rows correspond to Alice's unitaries and columns correspond to Bob's. Each measurement outcome denoted in an entry occurs with probability one under corresponding joint unitaries. Entries without a recorded value correspond with no recorded value. Unitaries correspond to input values as follows, $1 \iff U_i, 2 \iff U_i', 3 \iff U_i'', 4 \iff U_i'''$. Each recording strategy corresponds with $2x_a * x_b + 2 x_a + x_b \: \text{mod 4}$.}
    \label{fig:nonlinear}
\end{figure}

\section{Discussion}
The constructions presented above provide illustrative examples of functions that satisfy the security properties introduced in \cite{hartzell_quantum}. While this work focuses on functions with cardinality at most four, the general construction developed in Section 10 of \cite{hartzell_quantum} extends naturally to higher-dimensional systems, permitting the realization of substantially richer classes of functions who satisfy the properties of the general recording strategies. We leave the complete characterization of the functions that satisfy these security properties in arbitrary dimension for future research.

Beyond their enhanced security and efficiency properties in quantum key distribution applications, these examples demonstrate that parties may securely generate shared functions that non-degenerately depend on their joint information without requiring pre-shared randomness. Not only are these protocols resistant to ``harvest now, decrypt later'' attacks, but more broadly, they illustrate that quantum systems can enable parties to reach shared agreements even when the only physical record produced by the protocol contains no information about the agreement itself. This suggests a new role for distributed quantum systems as neutral arbiters of decentralized information-sharing, where these systems need not be trusted. 

\clearpage

\newpage

\printbibliography

\newpage

\section{Appendix}
\subsection{Secure Key Rates}\label{ref:secure_function_rates}
For protocols in which $P(\hat{s} = s | \tilde{Z}, j) = \frac{1}{2}$, under every iteration of the protocol, privacy amplification is necessary if there is concern over \textit{additional} side-channel attacks that could arise from methods other than prepare-and-measure attacks. When the chosen basis yields measurement outcomes approximately deterministically, error correction is only necessary to the extent to which the measurement device itself might be faulty, or if there is reason to believe that a non-optimal attack was deployed. 

As previously noted, in some protocols, at least one player may detect errors \textit{during} the protocol itself, and therefore may abort if errors are sufficiently high. 

\subsubsection{Per-Iteration Security Rates}

To underscore the differences in security guarantees between the proposed protocols and traditional \cite{BennettBrassard1984}-type implementations, we will focus on key rates derived from information leakage and error rates \textit{per iteration} of the protocol. This differs from existing asymptotic derivations as in  \cite{BennettBrassard1984}-type implementations, as these protocols do not have guarantees per iteration or conditional on attack. 


Following \cite{DevetakWinter2005}, the asymptotic function rate \textit{per protocol iteration} is given by\footnote{The standard formula is instead given per \textit{kept} round of the protocol \text{after} sifting and privacy amplification.}

 \[
    R = P_{\mathrm{keep}}\,[\,H(S|E) - h(P_{error})\,],
\]
where $S$ represents the random variable corresponding to the value $s \in \mathcal{A}$ shared by the legitimate parties, $s_a = s_b$ after measurement.

$E$ denotes the quantum system (or classical side information) accessible to a potential adversary.
$H(S|E)$ is the conditional von Neumann entropy of $S$ given $E$, which quantifies the residual uncertainty of the shared variable \(S\)  from the perspective of an observer holding arbitrary quantum information \(E\). 

$h(P_{error})$ is the binary Shannon entropy,
     \[
        h(Q) = -Q \log_2 Q - (1-Q)\log_2(1-Q),
    \]
    which quantifies the information cost of error correction per $s$.

\subsubsection{Key Rates Conditional on Attack}
We consider the above defined per iteration function rates, \textit{before} any privacy amplification is performed.  We may equivalently express the function rate conditioned on a (notional) measurement-attack flag.  Let \(H(S\mid E,\mathrm{no})\) denote the conditional von Neumann entropy of \(S\) given the adversary's system \(E\) in the \emph{no-attack} branch, and let \(h(P_{error, \mathrm{no}})\) denote the
binary entropy of the corresponding error rate \(P_{error, \mathrm{no}}\). These quantities include any effects of channel noise or device imperfections present during normal operation, in addition to the effects of non-measurement side-channel attacks. 
Similarly, let \(H(S\mid E,\mathrm{att})\) and \(h(P_{error, \mathrm{att}})\)
denote the analogous quantities in the \emph{attack} branch alone. The quantities represent the loss of secrecy and increase in error attributable \emph{solely} to a prepare-and-measure attack, independent of background noise or other system effects. Let
\[
p \coloneqq \Pr[\text{prepare-and-measure attack}]
\]
be the probability that a given round is subject to a prepare-and-measure attack (the flag is a notational device and need not be observable).

Then the overall key rate, including the retention probability
\(P_{\mathrm{keep}}\), can be written as the convex combination
\[
R \;=\; P_{\mathrm{keep}}\Big[
    (1-p)\bigl(H(S\mid E,\mathrm{no}) - h(P_{error, \mathrm{no}})\bigr)
    + p\bigl(H(S\mid E,\mathrm{att}) - h(P_{error, \mathrm{att}})\bigr)
\Big].
\]

In MDI-QKD protocols with (\cite{BennettBrassard1984}) states, $H(s\mid E,\mathrm{att}) - h(P_{error, \mathrm{att}}) = h(.75) - h(.25) = 0$, whereas in the proposed protocols, $H(S\mid E,\mathrm{att}) - h(P_{error, \mathrm{att}}) = 1$. 

$$R_{Proposed} \geq \frac{1}{2}\Big[
    (1-p)\bigl(H(S\mid E,\mathrm{no}) - h(P_{error, \mathrm{no}})\bigr)
    + p
\Big] $$

$$R_{MDI-QKD} = \frac{1}{4}\Big[
    (1-p)\bigl(H(S\mid E,\mathrm{no}) - h(P_{error, \mathrm{no}})\bigr)
\Big] $$

$(H(S\mid E,\mathrm{no}) - h(P_{error, \mathrm{no}})$ in the proposed protocols would need be significantly smaller than the MDI-QKD counterpart in order to produce an even approximately equal asymptotic function rate \textit{prior to privacy amplification}, with this difference scaling in $p > 0$.

Optimal error correction procedures will depend on the parties' objectives-- namely whether they wish to generate secure strings or single-iteration joint functions $s$. We leave the question of qualifying the optimal post-processing error correction procedures under single-iteration settings for further research.

\end{document}